# Physically transient photonics: random vs. distributed feedback lasing based on nanoimprinted DNA

*Andrea Camposeo,*[*,†] *Pompilio Del Carro,*[†] *Luana Persano,*[†] *Konrad Cyprych,*[‡] *Adam Szukalski,*[‡] *Lech Sznitko,*[‡] *Jaroslaw Mysliwiec,*[‡] *Dario Pisignano*[*,†,¶]

[†] National Nanotechnology Laboratory of Istituto Nanoscienze-CNR, via Arnesano I-73100, Lecce, Italy.

[‡] Institute of Physical and Theoretical Chemistry,Wroclaw University of Technology, Wyb. Wyspianskiego 27, 50-370 Wroclaw, Poland.

[¶] Dipartimento di Matematica e Fisica "Ennio De Giorgi", Università del Salento, via Arnesano I-73100, Lecce, Italy.

ABSTRACT. The authors report on a room-temperature nanoimprinted, DNA-based distributed feedback (DFB) laser operating at 605 nm. The laser is made of a pure DNA host matrix doped with gain dyes. At high excitation densities, the emission of the untextured dye-doped DNA films is characterized by a broad emission peak with an overall linewidth of 12 nm and superimposed narrow peaks, characteristic of random lasing. Moreover, direct patterning of the DNA films is demonstrated with a resolution down to 100 nm, enabling the realization of both surface-emitting and edge-emitting DFB lasers with a typical linewidth<0.3 nm. The resulting emission is polarized, with a ratio between the TE- and TM-polarized intensities exceeding 30. In addition, the nanopatterned devices dissolve in water within less than two minutes. These results demonstrate the possibility of realizing various physically transient nanophotonics and laser architectures, including random lasing and nanoimprinted devices, based on natural biopolymers.

Deoxyribonucleic acid (DNA) is gaining an increasing, prominent role as biomaterial for photonics and optoelectronics.[1,2] DNA solid state films are characterized by optical transparency in the 300-1,600 nm range, refractive index, $n$, of about 1.5 and low optical losses at telecom wavelengths. These properties have stimulated many research efforts aimed at exploiting the potentialities of such biopolymer for optical and electronic applications. Recent breakthroughs include the use of the DNA biopolymer as gate insulator in organic field-effect transistors,[3] as electron blocking layer in organic light-emitting diodes,[4] and as nonlinear optical material for second harmonic generation.[5] Furthermore, DNA-based waveguides[6,7] and lasers[8-10] have been realized. The latter application is particularly interesting since some lasing chromophores show enhanced emission in DNA compared to typically used synthetic polymer matrices.[8] Moreover, DNA is available as a waste product of the fishing industry and constitutes a renewable and biodegradable material for realizing "green" solid state biolasers,[11,12] which would allow the realization of physically transient, water-dissolvable photonic structures, complementing the recently-reported transient electronic devices.[13]

However, to date most of the reported DNA-based lasers have utilized surfactant-modified DNA,[14] which allows the complex DNA-cationic surfactant complexes to be soluble in organic solvents, whereas pure DNA is water soluble. Little is known about the exploitation of pure DNA in solid state lasers, although films of pure DNA doped with dyes have been demonstrated to show good gain coefficients.[15] In part, this is due to the lack of fully DNA-compatible patterning methods, which are needed to fabricate solid state laser devices. Patterning techniques that entirely work at room temperature are preferable in this respect, avoiding thermal-induced reversible DNA denaturation. For instance, denaturation is reported to be at 87.5 °C in 0.15 M sodium chloride with 0.015 M sodium citrate, for selected DNA origin, of about 2000 base pairs.[16] Nanopatterning at room temperature would therefore reduce possible DNA oxidation, as well as the degradation of the optical gain properties of guest chromophores. A few works have explored routes for patterning films of DNA-cationic surfactant complexes, by electron-beams,[17] self-assembling[18] holography,[19] and by substrate patterning/functionalization,[8] whereas there are no reports about nanopatterning of pure DNA films. To this aim, room-temperature nanoimprint lithography (RT-NIL), which has been demonstrated to be very effective for the production of lasers by conjugated polymer films and multilayers,[20] can represent a valuable choice for patterning biopolymers with high spatial resolution. In RT-NIL, irreversible deformations are induced in optical materials by pressing a rigid nanostructured template against the surface of the target films, namely without using high-temperature steps as in conventional NIL.[21]

Here we report on the realization of a distributed feedback (DFB) laser based on pure DNA doped with a laser dye, by an entirely green process involving the use of aqueous solutions and room-temperature nanopatterning. A clear transition from random lasing to DFB lasing is observed in films upon patterning. 1-dimensional gratings with feature resolution down to 100 nm are transferred to the films, and the resulting DFB devices show polarized laser emission at about 605 nm, with a threshold excitation density of few mJ cm$^{-2}$. The nanopatterned devices are demonstrated to be dissolvable in water within less than two minutes, opening interesting routes for the development of physically-transient organic photonic platforms.

RESULTS AND DISCUSSION

The chemical structure of DNA, Rhodamine 6G (R6G) and the scheme of Rhodamine binding are presented in Fig. 1. The emission spectra of pristine DNA films doped with R6G upon pulsed optical pumping are shown Fig. 2a,b. Above a pumping threshold fluence of about 7 mJ cm$^{-2}$, the broad photoluminescence (PL) spectrum (shown as a dashed line in Fig. 2a) collapses in a peak with linewidth of 12 nm and superimposed narrow lines (width < 0.3 nm, limited by the spectral resolution of the detection system), whose number and intensity increase upon increasing the excitation fluence (Fig. 2b). The presence of these complex emission features can be attributed to coherent random lasing, due to the scattering of light by the microscopic morphological and compositional inhomogeneities of the DNA-R6G blend, as observed also in dye-doped DNA-surfactant systems.[9,22] In order to better understand the origin of these characteristic emission peaks, we perform a Fourier analysis of an ensemble of individual emission spectra.[23] The ensemble-averaged amplitude of the Fourier transform (FT) of many emission spectra is shown in Fig. 2c, highlighting peaks spaced by a path length $\Delta d \cong 10$ μm, that corresponds to a characteristic cavity length of the order of 20 μm. These lengths are therefore attributed to intrinsic sample inhomogeneities (bright spots in the confocal fluorescence images of the inset of Fig. 2c) of the pristine DNA-R6G random lasing films. These domains have sub-micron sizes, and spatial separations well within the range 2-20 μm.

For some applications, laser emission at well determined wavelengths is preferred, a feature that requires the inclusion of an optical cavity in the active material. To this aim we realize DFB structures by a single step imprinting process on the DNA-based films. Figure 3a-d shows the surface topography of the imprinted dye-doped DNA samples, reproducing faithfully the master templates (Fig. 3a,b) and exhibiting well-defined features with height of 200 nm (Fig. 3c,d). The room-temperature imprinting parameters (pressing time = 5 minutes, applied force = 3 kN) on DNA are comparable to those optimized for other, thermoplastic polymeric matrices, such as poly(methylmethacrylate). The minimal feature size obtained on DNA is of the order of 100 nm, as shown in the inset of Fig. 3c, displaying the surface of a transferred grating with period Λ=200 nm.

The continuous wave (cw) PL spectra of the textured DNA-R6G films, collected at variable angles in a plane perpendicular to the imprinted grating grooves, are displayed in Fig. 3e. For each angle, the PL is enhanced at specific wavelengths, because of the out-coupling of the modes guided into the slab waveguide formed by the quartz substrate ($n$=1.46), the DNA film ($n$=1.55)[24] and air (the cw PL spectrum of untextured film is shown in Fig. 3f for comparison). Indeed, the imprinted grating allows part of the light guided within the DNA film to be scattered out along a specific forward direction, an effect that can be accounted for by considering the well-known Bragg relation related to the conservation of the in-plane components of the light wavevector:[25]

$$\cdot \ \frac{2\pi}{\lambda}\sin(\theta) = \pm\frac{2\pi n_{eff}}{\lambda} \pm m\frac{2\pi}{\Lambda} \qquad (1)$$

In this expression, $n_{eff}$ is the effective refractive index of the guided modes and $m$ is the diffraction order. This relation allows us to estimate the wavelength dependence of the effective refractive index (inset of Fig. 3f), which is in the range 1.50-1.56 in the investigated spectral range (550-650 nm) for the fundamental guided mode.

The presence of the imprinted grating is effective to suppress the random lasing emission and force the net amplification at determined wavelengths. The emission spectra of the DNA-based DFB devices upon pulsed optical pumping are shown in Fig. 4. For devices with a grating having Λ=400 nm, a narrow emission peak (linewidth<0.3 nm) at 605 nm appears for excitation fluences higher than 6 mJ cm$^{-2}$ (Fig. 4a), whose intensity increases linearly upon increasing the pumping fluence (Fig. 4b), a trend characteristic of lasing. Given the emission wavelength ($\lambda_{las}$=605 nm) and the effective refractive index ($n_{eff}$=1.54), this DFB structure is found to operate at the second order ($m$=2) according to the Bragg expression: $\lambda_{las} = 2n_{eff}\Lambda/m$ . First-order emission spectra of a DNA-based film imprinted with a grating having $\Lambda$=200 nm, are shown in the inset of Fig. 4b. The main difference between the DFB devices having different order is in their emission direction. The second-order and the first-order structures lase mainly along the direction normal to the film surface and from the device edge, respectively. In both systems, we observe the presence of two spectrally-resolved emission peaks (Fig. 4c) with a spectral distance of $\Delta\lambda\cong$0.5 nm, that is typical of DFB laser structures where index coupling is the dominant feedback mechanism.[26] In addition, the splitting between the two lasing modes is related to the effective length ($L_{eff}$) of the cavity:[26] $\Delta\lambda = \lambda_{las}^2/2L_{eff}$ . For the investigated devices we estimated an effective cavity length in the range 0.5-1 mm, that is slightly shorter than the length of our DFB devices ($L$>2 mm). This result evidences that only part of the active imprinted area contributes to the lasing action, a finding which is in agreement with the presence of $L_{eff}-$ limiting scattering centers in the DFB cavities, similar to those originating random lasing in pristine samples (Fig 2). We also find that the DFB emission is strongly polarized along a direction parallel to the grooves of the imprinted grating. Fig. 4c shows the lasing spectra collected through a polarization filter with axis parallel (TE mode) and perpendicular (TM mode) to the grooves, highlighting that the TE mode intensity is more than 30 times higher than the TM mode. Such polarization ratio is comparable to values found on highly-polarized DFB devices based on solution-processed conjugated polymers or thermal-evaporated molecules.[27]

An especially appealing feature of this platform to realize photonic devices, consist in the possibility of producing physically transient light-emitting components. Applications where biodegradable or transient, water-soluble devices may be advantageous include biophotonics, implantable analytic chips, and wireless optogenetics.[28] A collection of images showing the time sequence of dissolution of the nanopatterned lasers and of the DNA-R6G films upon immersion in DI water is displayed in Fig. 5. The pattern is rapidly dissolved, in a time interval of the order of seconds, and both the biopolymer and the light-emitting molecule gradually detach from the glass substrate underneath and diffuse into water. A complete passage from the solid state to solution is observed within two minutes, demonstrating the possibility of realizing transient photonic nanostructures by using pure DNA combined with suitable dyes or other light-emitting molecules. To the best of our knowledge, this is a first attempt to implement physically-transient photonics with organic lasers. Various intriguing developments can be envisaged based on the here reported results. The device dissolution time can be finely tailored upon embedding the pattern by means of protective, transparent layers. More interestingly, other classes of nanostructures, including light-emitting nanofibers based on dye-doped DNA, could provide alternative material systems and allow new lasing architectures to be explored. These experiments are currently in progress in our laboratories.

CONCLUSIONS

In summary, surface and edge-emitting DFB laser structures based on a pure DNA matrix doped with a laser dye has been reported. The DFB devices are based on nanostructured DNA films, patterned by RT-NIL. The films show random lasing features in their pristine form, which clearly moves to DFB lasing following nanopatterning. The realized DFB devices exhibit polarized emission at 605 nm, with a linewidth of 0.3 nm and a threshold excitation density of 6 mJ cm$^{-2}$. The proposed method and devices pave the wave for the development of biocompatible and biodegradable nanophotonics, through fully green approaches combining the use of aqueous solutions with environmentally friendly, room-temperature lithographies for material processing.

In perspective, these laser structures can be utilized for designing novel schemes for cell stimulations and/or imaging, and intriguing information-communication technology routes using solid state DNA as simultaneous carrier of both genetic and photonic information.

EXPERIMENTAL SECTION

Samples are prepared using salmon – Oncerhynchusketa DNA derived from sperm and tastes, purchased from Sigma-Aldrich as a sodium salt, with 41.2 % content of Guanine-Cytosine base pairs. The average length of DNA strands, has been previously estimated to around 2000 base pairs by the group of Tanaka.[29] Films with thicknesses in the range 0.7-2.5 µm are produced by spincoating (500-700 rpm) onto quartz substrates (1×1 cm$^2$) a water solution of DNA and R6G, with a 1 % of R6G:DNA weight ratio. The great advantage of DNA is its ability to order the dye molecules in such a way that self-quenching effects are prevented. Such hypothesis has been drawn by the group of Kawabe and assumes the process of grove binding or dye intercalation.[30] Films are then dried in air at 20°C and at about 40% of relative humidity, for about two days before imprinting. Gratings with periods Λ=200 nm and Λ=400 nm are realized by RT-NIL using Si master templates produced by electron-beam lithography and reactive ion etching.[20,31] The patterned surface of the master templates is placed in contact with the free surface of the DNA films and a pressure of 100 MPa is then applied by means of a two-column precision manual press. Following master detachment and grating transfer, the morphology of the structured films is investigated by atomic force microscopy, using a Multimode system (Veeco) equipped with a Nanoscope IIIa controller and working in Tapping mode by Si cantilevers. The emission of the nanostructured films is firstly characterized by angle-resolved cw PL spectroscopy, exciting samples by a diode laser ($\lambda_{exc}$=405 nm) and collecting the PL signal by an optical fiber mounted on a rotating stage and analyzed by a monochromator (USB4000, Ocean Optics). For characterizing the lasing emission, samples are excited by the second harmonic of a Q-switched neodymium doped yttrium aluminum garnet pulsed laser (10 Hz repetition rate), with polarization parallel to the DFB grating grooves and spot size of $2.0 \times 0.4$ mm$^2$. The emission is collected by a spherical lens ($f$=75 mm) and coupled to an optical fiber and a monochromator (iHR320, Jobin Yvon) equipped with a charge coupled device (CCD) detector and a 1200 grooves/mm grating for spectral investigation. Samples are held in a vacuum chamber ($10^{-4}$ mbar) during measurements.

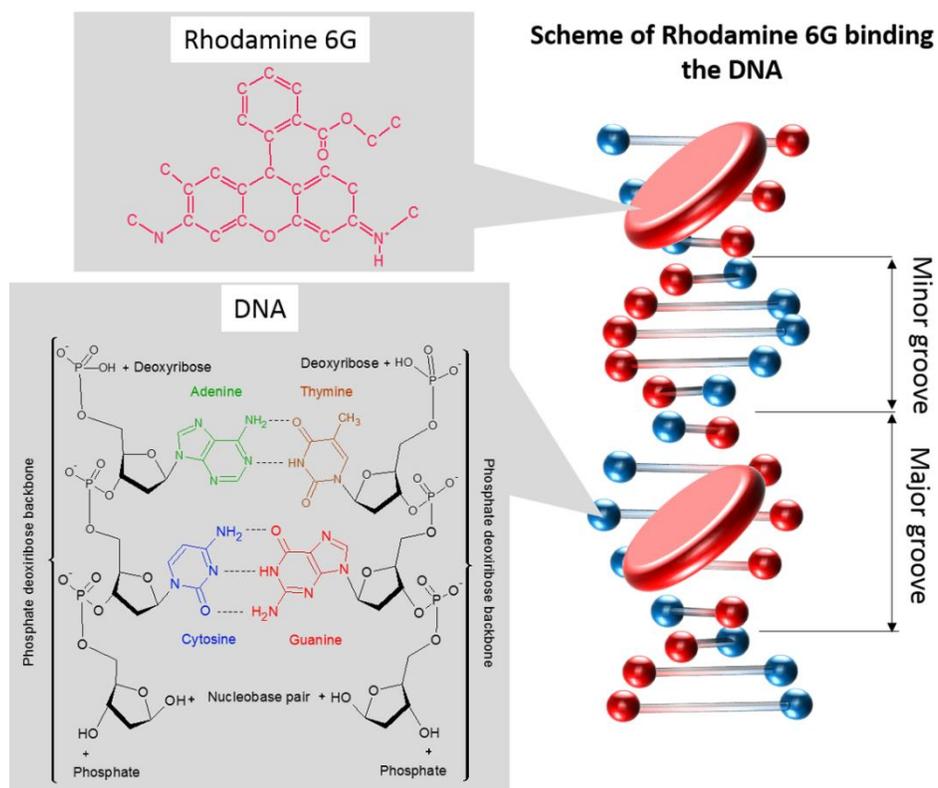

**Figure 1.** Chemical structure of DNA and R6G dye (left), and schematic illustration of R6G binding to DNA helix (right).

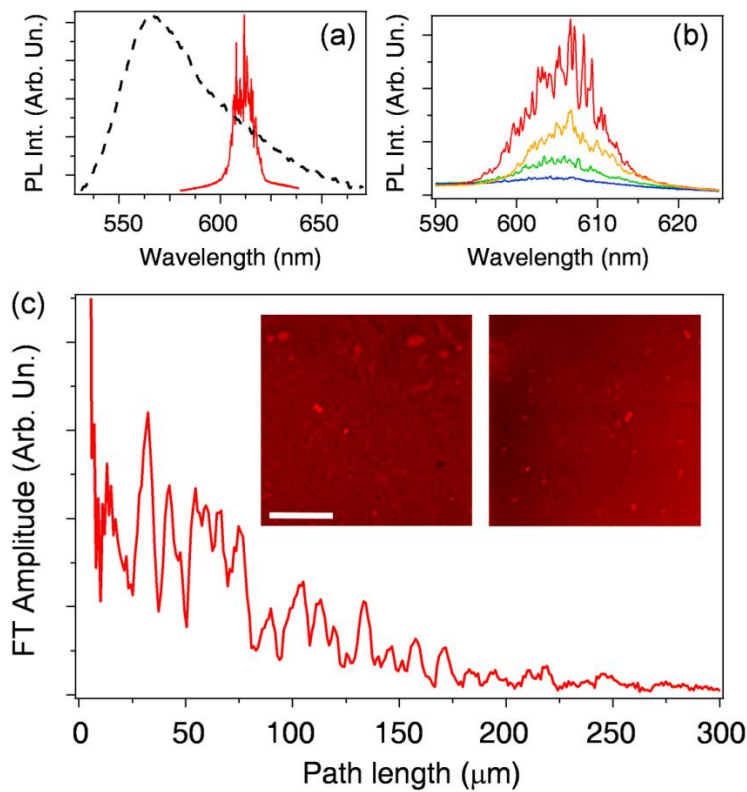

**Figure 2**. (a) PL spectra of a pristine DNA-R6G film upon cw excitation (dashed line) and pulsed excitation at a fluence of 15 mJ cm$^{-2}$ (continuous line). (b) Single shot emission spectra collected at different pumping fluences. From bottom to top: 8 mJ cm$^{-2}$, 10 mJ cm$^{-2}$, 12 mJ cm$^{-2}$ and 15 mJ cm$^{-2}$, respectively. (c) Amplitude of the FT of the emission spectra from the dye-doped DNA films. Data are calculated by averaging the FT spectra of individual excitation pulses. Inset: fluorescence confocal micrographs of DNA-R6G films. Scale bar: 10 μm.

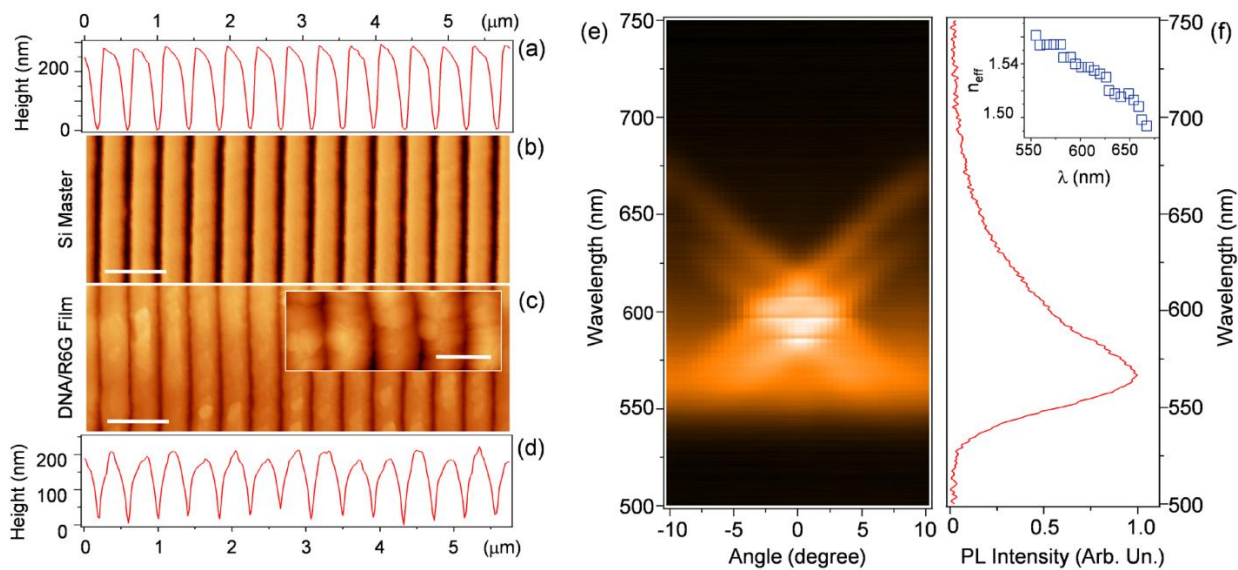

**Figure 3**. (a) Atomic force microcopy (AFM) topographic images of the Si master template (Λ=400 nm) and of DNA-R6G films textured by RT-NIL. Height profiles (a, d) and AFM micrographs (b, c) of the Si master (a, b) and of the imprinted DNA-R6G films (c, d), respectively. Scalebar in (b) and (c): 800 nm. Inset in (c): AFM micrograph of a DNA-R6G film patterned with a grating period, Λ=200 nm. Scale bar: 250 nm. (e) Map of the angle-dependent PL of a patterned DNA-R6G film with period Λ=400 nm. The bright regions are associated to the optical modes out-coupled by Bragg-scattering. (f) Cw PL spectrum of an untextured DNA-R6G film. Inset: spectral dispersion of the effective refractive index, $n_{eff}$.

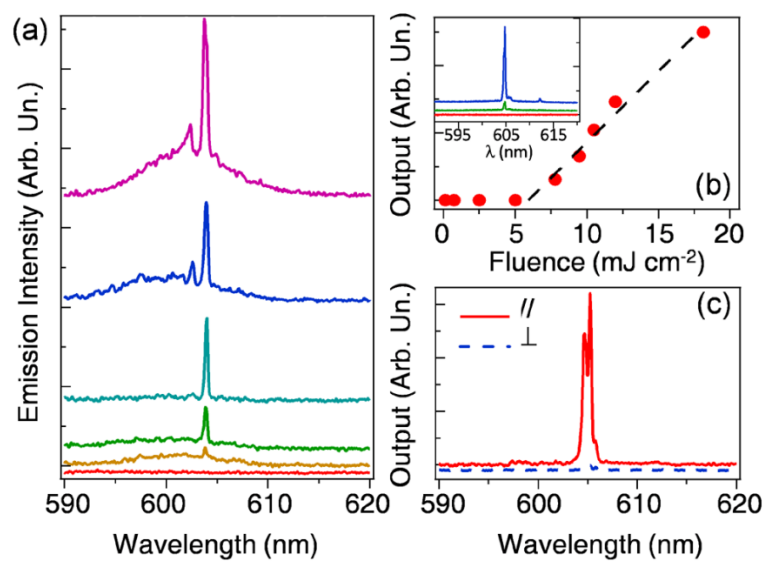

**Figure 4**. (a) Lasing spectra from a DNA-R6G DFB structure with Λ=400 nm. From bottom to top, excitation fluence: 0.7 mJ cm$^{-2}$, 7.8 mJ cm$^{-2}$, 9.5 mJ cm$^{-2}$, 10.5 mJ cm$^{-2}$, 12 mJ cm$^{-2}$ and 18 mJ cm$^{-2}$. Spectra are vertically shifted for better clarity. (b) Lasing intensity vs. pumping fluence for a second order DFB DNA-based structure. Inset: lasing spectra from a DNA-R6G structure with a grating period of Λ=200 nm. From bottom to top: excitation fluences: 5 mJ cm$^{-2}$, 10 mJ cm$^{-2}$, 18 mJ cm$^{-2}$. (c) Polarized lasing spectra (Λ=400 nm): polarization components parallel (continuous line) and perpendicular (dashed line) to the imprinted grating lines. Excitation fluence = 9 mJ cm$^{-2}$.

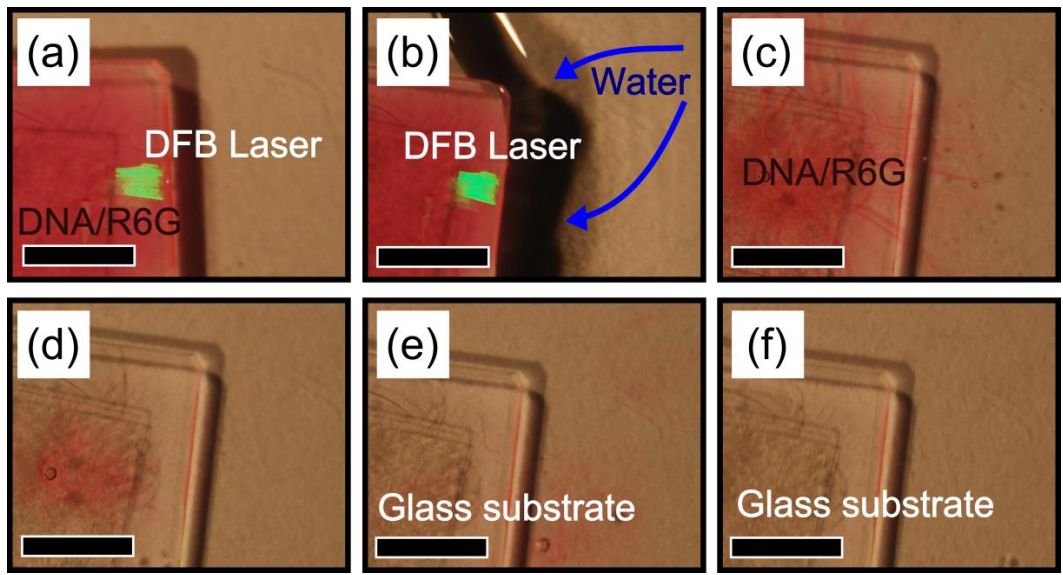

**Figure 5**. Optical images showing the dissolution of the DNA-R6G DFB structures in water. (a) Reddish DNA-R6G, with the nanopatterned region constituting the DFB laser imaged in green due to light diffraction. (b) Photograph of the instant at which the device is immersed in water ($t$ = 0). (c) $t$ = 30 s. The pattern rapidly dissolves in water, and the DNA-R6G molecules diffuse into the liquid. (d) $t$ = 60 s. Device dissolution continues. (e) $t$ = 90 s. The active film is almost completely dissolved, and the glass substrate underneath is now directly exposed to water. (f) $t$ = 120 s. (a-f): Scale bar: 5 mm.

AUTHOR INFORMATION

**Corresponding Author**

* Authors to whom correspondence should be addressed: A.C.: andrea.camposeo@nano.cnr.it;

D.P.: dario.pisignano@unisalento.it.

ACKNOWLEDGMENT

The research leading to these results has received funding from the European Research Council under the European Union's Seventh Framework Programme (FP/2007-2013)/ERC Grant Agreement n. 306357 (ERC Starting Grant "NANO-JETS").